\documentclass[amsmath,amssymb,reprint,aps,prl,showpacs,superscriptaddress,floatfix]{revtex4-1}

\usepackage[utf8]{inputenc}
\usepackage{graphicx}
\usepackage{color}
\usepackage{epstopdf}
\usepackage{braket}
\usepackage{siunitx}
\usepackage[normalem]{ulem}

\begin{document}

\title{Scaling of Yu-Shiba-Rusinov energies in the weak-coupling Kondo regime}
\author{Nino Hatter}
\author{Benjamin W. Heinrich}
\email{bheinrich@physik.fu-berlin.de}
\author{Daniela Rolf}
\author{Katharina J. Franke}
\affiliation{Fachbereich Physik, Freie Universit\"at Berlin,
                 Arnimallee 14, 14195 Berlin, Germany}

\date{\today}

\newcommand{\NH}{NH$_3$}
\newcommand{\didv}{$dI/dV$}
\begin{abstract}
The competition of the \textit{free spin }state of a paramagnetic impurity on a superconductor with its \textit{screened} counterpart is  characterized by the energy scale of Kondo screening $k_\mathrm{B} T_\mathrm{K}$ compared to the superconducting pairing energy $\Delta$.
When the experimental temperature suppresses Kondo screening, but preserves superconductivity, i.e., when $\Delta/k_\mathrm{B}>T>T_\mathrm{K}$, this description fails. Here, we explore this temperature range in a set of manganese phthalocyanine molecules decorated with ammonia (MnPc-\NH)  on Pb(111). We show that these molecules suffice the required energy conditions by exhibiting weak-coupling Kondo resonances. 
We correlate the energy of the Yu-Shiba-Rusinov (YSR) bound states inside the superconducting energy gap with the intensity of the Kondo resonance. We reveal that the bound state energy follows the expectations for a classical spin on a superconductor. This finding is important in view of many theoretical predictions using a classical spin model, in particular for the description of Majorana bound states in magnetic nanostructures on superconducting substrates.
\end{abstract}

\maketitle

A single magnetic atom/molecule adsorbed on a superconductor presents a local perturbation for the quasiparticles with a Coulomb and an exchange scattering potential. The exchange coupling $J$ leads to the formation of localized bound states inside the superconducting energy gap $\Delta$. These so-called Yu-Shiba-Rusinov (YSR) \cite{Yu1965,Shiba1968,rusinov_a.i._theory_1969} states can be detected by tunneling spectroscopy as a pair of resonances symmetrically around the Fermi level ($E_\mathrm{F}$)~\cite{yazdani_probing_1997,Ji2008,Heinrich2017rev}. The simplest description of the scattering assumes a classical spin $S$, 
 where the  bound-state energy $\varepsilon$ then depends on the coupling strength $J\rho_0$ ($\rho_0$ is the DoS at $E_\mathrm{F}$  in the normal state).  However, in many cases, a classical description is insufficient. The quantum mechanical nature of the spin manifests in a Kondo resonance outside the superconducting energy gap and in the normal state. Because both YSR and Kondo states are a result of the same exchange coupling strength $J\rho_0$, their energies are connected with each other by a universal relation \cite{Matsuura77,Satori1992,Sakai1993}. The formation of the Kondo singlet with its energy scale given by $k_\mathrm{B}T_\mathrm{K}$ (with $k_\mathrm{B}$ being the Boltzmann constant and $T_\mathrm{K}$ the Kondo temperature) thereby competes with the singlet state of the superconductor. 
A quantum phase transition from an unscreened,  {\it free spin} ground state to a {\it Kondo screened} state occurs at $k_\mathrm{B}T_\mathrm{K} \sim \Delta$.
Recent experiments corroborated the theoretically predicted universal relation between $\varepsilon$ and  $T_\mathrm{K}$ \cite{ Deacon2010b,Franke2011,Lee2014,Jellinggaard2016,Lee2017}. 

 An intriguing situation arises if the Kondo energy scale is ill-defined. This is the case when the thermal energy is larger than the energy scale of the Kondo screening. Then, the exchange coupling $J$ gives rise to scattering processes, which induce a  zero-bias resonance in transport measurements/tunneling experiments. The width of the resonance is only given by the experimental temperature, i.e., it is independent of $J$. 
The  scattering processes can be well captured within perturbation theory. This description is commonly referred to as weak-coupling Kondo \cite{Zhang2013,Ternes2016}. Contrary to the temperature-dependent Kondo description, the energy of the YSR bound state is temperature independent. The relation between YSR bound-state energy $\varepsilon$ and observables of the weak-coupling Kondo resonance has not been established to date.

\begin{figure}[b]
	\centering
		\includegraphics[width=0.48\textwidth]{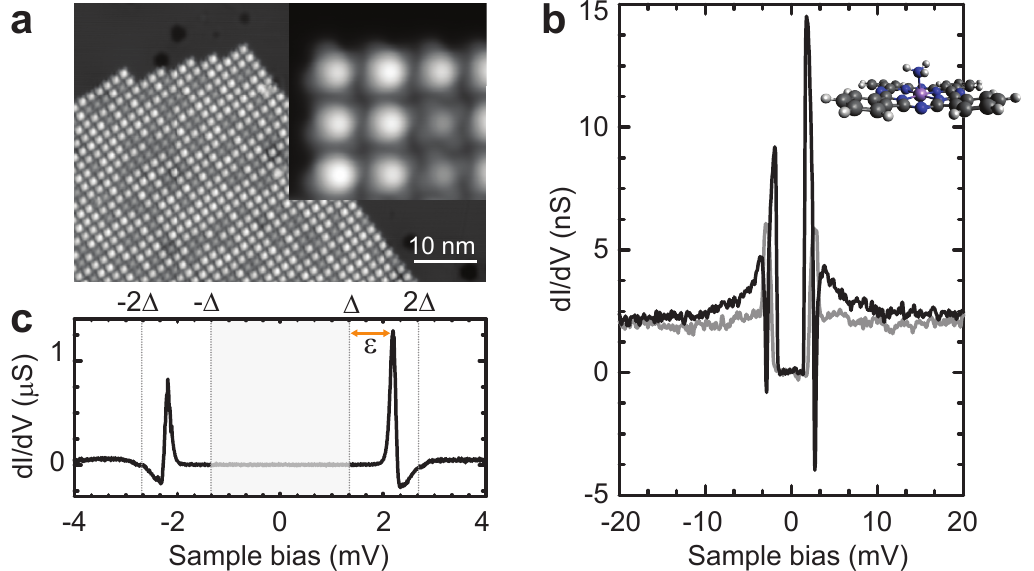}
	\caption{(a) Topography of a mixed island of MnPc-\NH\ and MnPc molecules (setpoint: $U=50\,$mV, $I=100\,$pA). The inset shows a zoom on 7 MnPc-\NH\ and two MnPc molecules ($50\,$mV, $200\,$pA). 
(b) Characteristic \didv spectrum of an MnPc-\NH\ (black) and on the bare Pb(111) (gray) acquired with a superconducting Pb tip ($50\,$mV, $200\,$pA, $U_{mod}=500\,\mu$V$_{rms}$). (c) Zoom on the superconducting gap presenting a pair of YSR resonances ($5\,$mV, $200$\,pA, $20\,\mu$V$_\mathrm{rms}$). The bound-state energy is indicated by $\varepsilon$ and the tip gap by the gray area.}
\label{Topo}
\end{figure}

Here, we experimentally deduce a new expression of the universal relation between exchange scattering processes in the weak-coupling Kondo regime with the bound-state energy $\varepsilon$ of the YSR states. We show how the height $\alpha$ of the zero-bias resonance correlates with the binding energy $\varepsilon$. 
The unraveling of this correlation demands for an ensemble with a variety of coupling strengths $J$, all being in the weak-coupling Kondo regime. An ideal system is a Moir\'e pattern of adsorbates on a superconductor, where each adsorbate bears a slightly different exchange potential for the substrates'  quasiparticles. Manganese phthalocyanine (MnPc) molecules on a Pb(111) surface exhibit such a Moir\'e pattern with a strong variation in $J$, but the scattering is not in the weak-coupling Kondo regime \cite{Shuai2010,Franke2011,hatter2015}. 
We use MnPc/Pb(111) as a template  and attach an additional \NH\ ligand to the Mn ion, borrowing a successful strategy from surface chemistry. The surface trans-effect reduces the coupling to the substrate \cite{Flechtner2007,Hieringer2011} while the variety of adsorption sites is maintained. We show that all these molecules exhibit a weak-coupling Kondo resonance and YSR states, where the height of the Kondo resonance and the YSR energy are connected via the exchange coupling strength $J\rho_0$.

\section{Results}
\subsection{Topographic appearance and characterization of MnPc-\NH}

Figure\,\ref{Topo}a shows a typical scanning tunneling microscopy (STM) topography of an MnPc island after \NH\ adsorption. While the square-like structure of the MnPc monolayer island is preserved, the appearance of the molecules is altered compared to the characteristic clover-shape pattern of the pristine MnPc molecule (see inset of Fig.\,\ref{Topo}a for a zoom on seven MnPc-NH$_3$ and two MnPc molecules). MnPc-\NH\ molecules appear nearly circular symmetric and $3.1\pm0.2$\,\AA\ high, i.e., $\approx1.7\pm$0.2$\,$\AA\ higher than MnPc. We interpret this in terms of \NH\ binding to the central Mn ion and protruding towards the vacuum. By means of voltage pulses we can desorb the \NH\ from the molecule in a controlled manner  and restore pristine MnPc (see Supplementary Information).

\subsection{YSR bound states as  indicator of magnetic coupling strength}
To study the exchange coupling strength between the molecules and the superconductor, we perform \didv\  spectroscopy with a superconducting tip (see Methods section for details). Figure\,\ref{Topo}b (black curve) shows a characteristic example of a \didv spectrum on MnPc-\NH. The spectrum is dominated by the superconducting gap structure in the range of $\approx\pm 3$\,meV around the Fermi level $E_\mathrm{F}$. At energies outside the superconducting gap, an increasing intensity is observed when approaching $E_\mathrm{F}$ from both bias sides. 
In the later part of this manuscript, we will show that this can be associated with a zero-bias resonance due to Kondo scattering.

Within the superconducting gap, a single pair of YSR resonances is observed (high-resolution zoom in Fig.\,\ref{Topo}c). The electron-like component,  i.e., the resonance at positive energy, is more intense than the hole-like resonance. This holds true for all MnPc-\NH\ molecules studied (see Supplementary Fig.~2d). We have studied a set of 44  molecules before and after controlled, tip-induced desorption of the \NH\ ligand. The comparison allows us to assign the bound-state energy $\varepsilon$ to the intense electron-like excitation, i.e., $\varepsilon >0$. 
Hence, all MnPc-\NH\ molecules are in the unscreened,  {\it free spin} ground state (see Supplementary Information for details). 

Variations in $\varepsilon$ for different molecules are observed, but $\varepsilon$ is always more positive than in the case of pristine MnPc. The small variations in $\varepsilon$ correlate with the stronger variations in $\varepsilon$ of the pristine MnPc molecules, i.e., after \NH\ desorption. Hence, the coupling strength to the substrate electrons $J$ again varies within the Moir\'e pattern, but is reduced compared to MnPc as expected for the surface trans-effect~\cite{Flechtner2007,Hieringer2011}. The addition of an axial ligand on the opposite side of the central ion pulls the Mn away from the surface and weakens the Mn--surface coupling.

\begin{figure}[tb]
	\centering
		\includegraphics[width=0.48\textwidth]{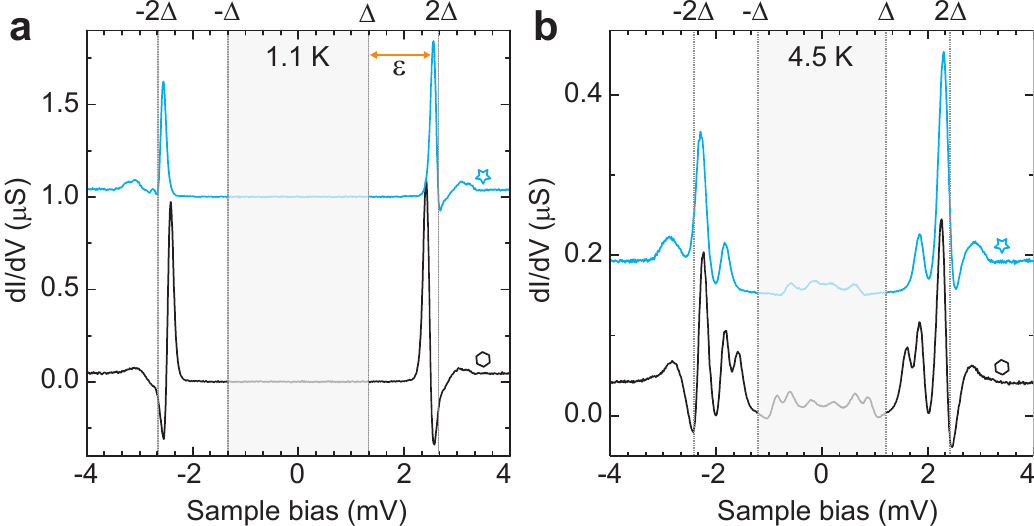}
	\caption{Two typical spectra of MnPc-\NH\ at 1.1\,K (a) and 4.5\,K (b), respectively, acquired with a superconducting Pb tip ($5\,$mV, 200\,pA, $20\,\mu$V$_{rms}$). Gray shaded areas indicate sample biases $|eV|\leq \Delta_{tip}$. At $4.5$\,K (b), additional resonances are observed because of tunneling out off/ into thermally excited YSR states.
}
\label{Shiba:4K}
\end{figure}

Unambiguous evidence for the {\it free spin} ground state is found by measurements at slightly elevated temperature. While at 1.1\,K, we observe only a single pair of YSR resonances for all MnPc-\NH\ molecules, we observe two or three pairs of resonances together with their thermal replica for all complexes at  4.5\,K (see two characteristic examples in Fig.~\ref{Shiba:4K}b). The separation of these resonances amounts to $200-400\,\mu$eV. In most cases, this splitting is  larger than the splitting of the resonance in the case of pristine MnPc. The absence of these resonance at lower temperature shows that they are linked to a thermally activated occupation of low-lying excited states. This behavior signifies an anisotropy-split ground state~\cite{hatter2015}, which can be explained by a  {\it free spin} $S=1$ with axial and transversal anisotropy \cite{Zitko2011}.
 We emphasize that a splitting of the excited YSR state would result in a set of three YSR pairs independent of the experimental temperature \cite{hatter2015}.
Note that, unlike in the case of iron phthalocyanine on Au(111), where \NH\ adsorption induces a spin change from an intermediate spin state of $S=1$ to a low spin of $S=0$~\cite{isvoranu_ammonia_2011}, we do not find an indication for a change of the spin state here.

\begin{figure}[tb] 
	\centering
		\includegraphics[width=0.48\textwidth]{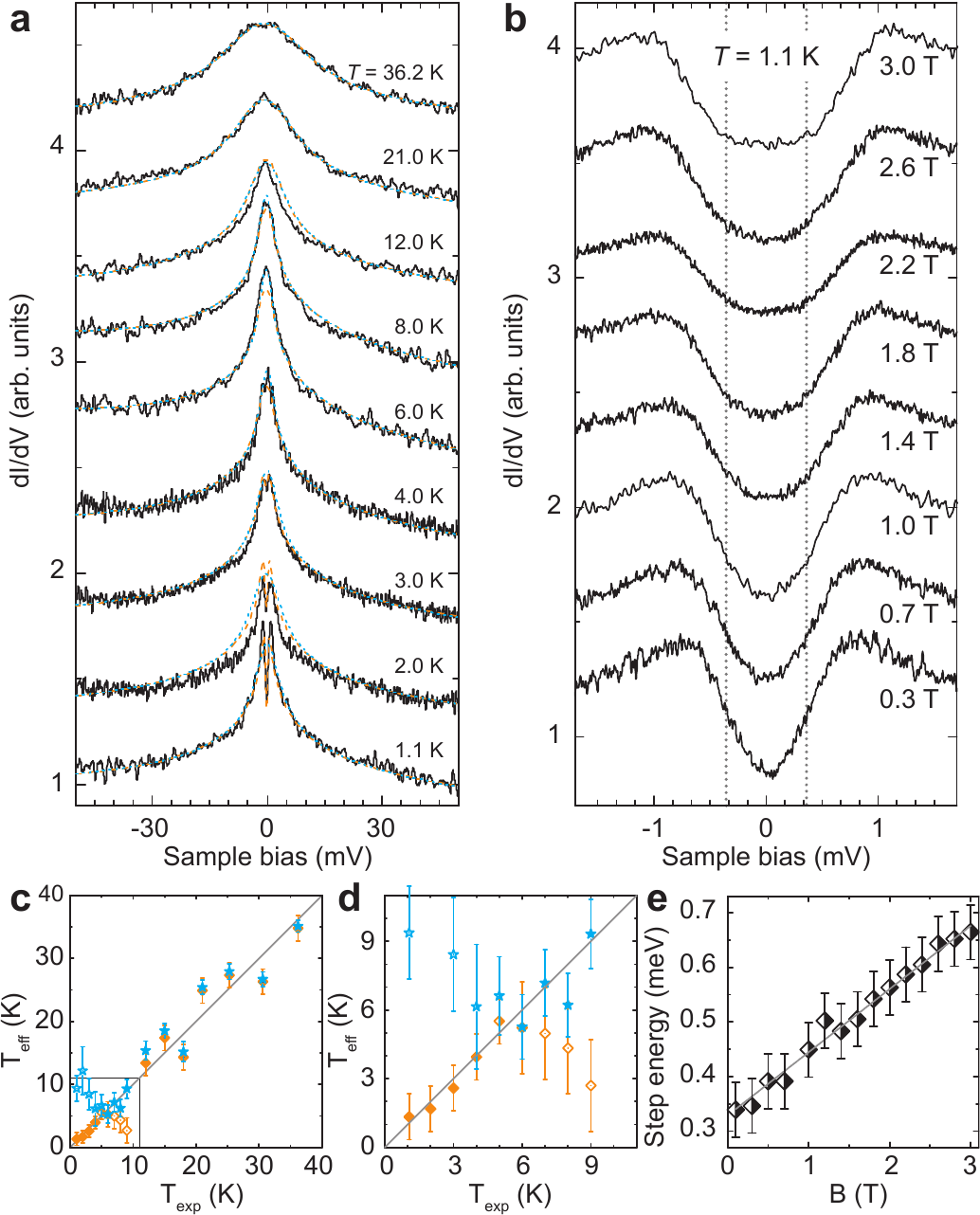} 
	\caption{$T$ and $B$ field dependence of the Kondo resonance.
(a) Temperature evolution of the zero-energy resonance at $B=0.1$\,T, which quenches superconductivity in the sample ($50\,$mV, 200\,pA, $500\,\mu$V$_{rms}$). Spectra are offset for clarity. A split zero-energy resonance is observed. Orange (blue) lines are fits according to the third order scattering model as described in Ref.~\cite{Zhang2013,Ternes2016} including (excluding) uniaxial magnetic anisotropy as determined in (e). 
(b) Evolution of the step-like feature around $E_\mathrm{F}$ in an external $B$ field parallel to the surface normal ($5\,$mV, 200\,pA, $50\,\mu$V$_{rms}$). 
(c,d) Effective temperature $T_\mathrm{eff}$ as extracted from fits like in (a) versus the experimental temperature $T_\mathrm{exp}$. Orange squares (blue stars) are from fits accounting for (disregarding) the axial anisotropy as determined in (e). 
(e) step energies versus $B$. Energies are extracted from the \didv spectra in (b) by a fit with symmetric step functions. The gray line indicates the fit to a Spin Hamiltonian for $S=1$ with uniaxial anisotropy: $D= -0.33\pm0.01\,$meV; $g=2.03\pm0.07$. All spectra are acquired with a normal metal Au tip. Error bars indicate the fit uncertainty. 
 }
\label{Kondo:B}
\end{figure}

\subsection{Kondo effect in the weak-coupling regime}
Next, we analyze the resonance outside the superconducting gap, which was already identified in Fig.\,\ref{Topo}b. For this, we quench the superconducting state of the substrate by applying a magnetic field of $B =0.1\,$T perpendicular to the sample surface and employ a normal metal tip.
Figure~\ref{Kondo:B}a shows typical spectra acquired on MnPc-\NH\  at various temperatures when the substrate is in the normal state. A zero-bias resonance is observed, which is reminiscent of the Kondo effect. At low temperatures, the peak is split around $E_{\mathrm{F}}$. 

We first focus on this splitting, which cannot be explained in terms of a Zeeman splitting, because the field strength is low compared to temperature~\cite{note1}.
To explore a possible magnetic origin of the splitting, we record \didv\ spectra on one molecule in fields up to $3\,$T. While the overall shape of the zero-bias resonance does not change (Supplementary Fig.~3), a zoom on  the dip at $E_\mathrm{F}$ unveils an opening of the gap with increasing out-of-plane field (Fig.~\ref{Kondo:B}b). Hence, the gap can be associated to inelastic spin excitations. The extracted step energies  (Fig.~\ref{Kondo:B}e) are fit to a simple Spin Hamiltonian, which assumes the anisotropy axis being parallel to the out-of-plane field: $\mathbf{\hat{\cal H}}=DS_z^2-g\mu_\mathrm{B} B_z\cdot S_z$. Here, $D$, $S_z$, $B_z$, $\mu_\mathrm{B}$, and $g$ are the axial anisotropy parameter, the projection  of the spin and the magnetic field  in $z$ direction, the Bohr magneton, and the  Land\'e g-factor, respectively. For a spin of $S=1$, D amounts to $-0.33\pm0.01$\,meV and $g$ is $2.0\pm0.1$.  

It is noteworthy that, on other MnPc-\NH\ molecules, we observed a linear dependence of the step energies on field strength only above $\approx 1$\,T and a slower increase at lower fields.  In these cases, the B-field dependence is well reproduced when accounting for an additional in-plane anisotropy term $E(S_x^2-S_y^2)$ in the Hamiltonian (see Supplementary Fig.~5). Occasionally, we also observe two pairs of excitation steps as is expected in the case of non-zero $E$ (Supplementary Fig.~4). Yet, most of the times, this is blurred by the limited energy resolution ($\approx 300\,\mu$eV at 1.1\,K). 

The above made observations of inelastic spin excitations fit to an $S=1$ spin system with dominant axial anisotropy and are in line with the measurements in the superconducting state. The YSR resonances are split into two (three) resonances separated by up to  400\,$\mu$eV, when measured at 4.5\,K, which indicates $E\approx 0$ ($E\neq0$).

Interestingly, the zero-field splitting is observed on top of a zero-bias resonance with a half width at half maximum (HWHM) of 3.7\,meV. In order to split a (strong-coupling) Kondo resonance of this width by means of an external magnetic field, a critical field of $B_\mathrm{c}\approx 65\,$T would be needed. Yet, the axial anisotropy of 0.3\,meV -- this is equivalent in energy to a field of 3\,T -- is sufficient to induce a sizable splitting. The absence of a critical field contradicts an explanation of the zero-bias resonance as an expression of a strong-coupling Kondo effect. However, the zero-bias resonance agrees with the system being in the weak-coupling Kondo regime.

To corroborate this interpretation, we performed temperature-dependent measurements (Fig.~\ref{Kondo:B}a). With increasing temperature, the height of the resonance decreases and its width increases. The symmetric steps close to the Fermi level broaden and vanish at $T\approx 4\,$K. 
Following our arguments derived from the B-field dependence, the zero-bias resonance in the \didv\ spectra shall be described by scattering at the impurity spin during tunneling \cite{note2}. The formalism of the weak-coupling Kondo effect has been 
described by Anderson and Appelbaum~\cite{Appelbaum1966,Anderson1966,Appelbaum1967} within second order perturbation theory, which account for third order scattering processes. To test this scenario, we fit the spectra using the scattering approach as described in Ref.~\cite{Ternes2016}, accounting for the axial anisotropy as determined above (orange lines in Fig.~\ref{Kondo:B}a). The fit accounts for a broadening of the logarithmic zero-bias divergence by an effective temperature  parameter $T_{\mathrm {eff}}$. In Fig.~\ref{Kondo:B}c we draw $T_{\mathrm {eff}}$ against the experimental temperature $T_{\mathrm {exp}}$. 
In a broad range of $T_{\mathrm {exp}}$, the values fall onto the identity line (orange squares in Fig.~\ref{Kondo:B}c), which is good evidence for the weak-coupling Kondo regime. 

We note that in the temperature range of $6$ to $9$\,K, $T_{\mathrm {eff}}$ drops below  $T_{\mathrm {exp}}$. Interestingly, the deviation starts  when the thermal energy $k_\mathrm{B}T$ surpasses the anisotropy energy of 330$\,\mu$eV. Apparently, the thermal scattering diminishes the effect of the magnetocrystalline anisotropy. We fit the \didv\ spectra again with the scattering approach, but this time without any anisotropy term (blue stars). Then, $T_{\mathrm {eff}}$ is larger than $T_{\mathrm {exp}}$ below 4\,K, but falls on the identity line above. 
Hence, these experiments show how magnetocrystalline anisotropy can affect third order scattering signatures, but that the effect is diminished by temperature. 

\begin{figure}[tb] 
	\centering
		\includegraphics[width=0.48\textwidth]{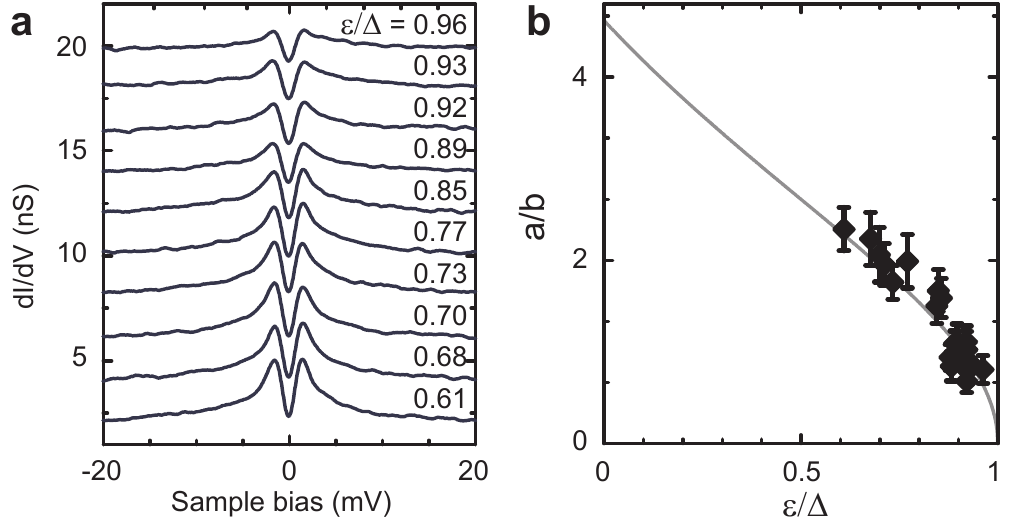} 
	\caption{ Kondo scattering vs. YSR energy. (a) Selected \didv\ spectra of different MnPc-\NH\ complexes ordered according to the YSR energy ($50\,$mV, 200\,pA, $500\,\mu$V$_{rms}$).  Spectra are acquired at $B=2.7\,$T in order to quench superconductivity in the Pb covered tip and the sample. Spectra offset for clarity. (b) Amplitude over background $a/b$ as a function of $\varepsilon/\Delta$. $a/b$ extracted from spectra as in (a). $\varepsilon/\Delta$ as extracted from spectra at $B=0\,$T similar to Fig.~\ref{Topo}c.  Error bars indicate the uncertainty of the read out. For $\varepsilon$ the error is in the order of the symbol size. }
\label{Kondo:YSR}
\end{figure}

 \subsection{Correlation of weak-coupling Kondo resonance with YSR energy}
Next, we attempt to correlate both expressions of scattering at the magnetic adsorbate, i.e., YSR and weak-coupling Kondo physics, in order to establish a universal picture.
Both, the third order scattering and the YSR bound states depend on the exchange coupling strength $J\rho_0$. 
Assuming a classical spin, the bound-state energy $\varepsilon$ was predicted to scale as $\varepsilon/\Delta= (1-\alpha^2)/(1+\alpha^2)$, with $\alpha=\pi S J\rho_0$~\cite{Shiba1968,rusinov_a.i._theory_1969}.

Using second order perturbation theory, the amplitude $a$ of the zero energy resonance caused by third  order spin scattering scales with $(J\rho_0)^3$~\cite{Zhang2013}.  The apparent height of the molecule of 3.1\,\AA\ prohibits direct tunneling to the substrate. Hence, we can assume the background conductance $b$ to be mainly caused by second order spin scattering (elastic and inelastic) and scaling with $(J\rho_0)^2$.
Then we can approximate: 
\begin{equation}
\frac{a}{b} \propto \frac{1}{\pi S} \, \sqrt{\frac{1-\varepsilon/\Delta}{1+\varepsilon/\Delta}}\,.
\label{Eq:amplitude}
\end{equation}
The variations in $J\rho_0$ caused by the different adsorption sites of the molecules within the Moir\'e allow us to test this relation. We first determine $\varepsilon$  of different complexes using a superconducting tip.  We then measure \didv\ spectra of the zero bias resonance on the same molecules at $B=2.7$\,T \cite{note3} and extract the amplitude $a$. Figure~\ref{Kondo:YSR}a shows a selection of such spectra ordered according to their YSR energy $\varepsilon$.  With increasing $\varepsilon$, the amplitude $a$ decreases. In Fig.~\ref{Kondo:YSR}b, we plot the  amplitude-over-background ratio $a/b$ as a function of YSR energy $\varepsilon$ over $\Delta$. A fit to Eq.~\ref{Eq:amplitude} with a linear scaling as the only free parameter is shown in gray and  describes the data evolution well. 
Hence, we conclude that the description of a YSR impurity as a classical spin is a sufficient model in the case of the Kondo effect in the weak-coupling regime. 

\section{Discussion}

\begin{figure}[t] 
	\centering
		\includegraphics[width=0.48\textwidth]{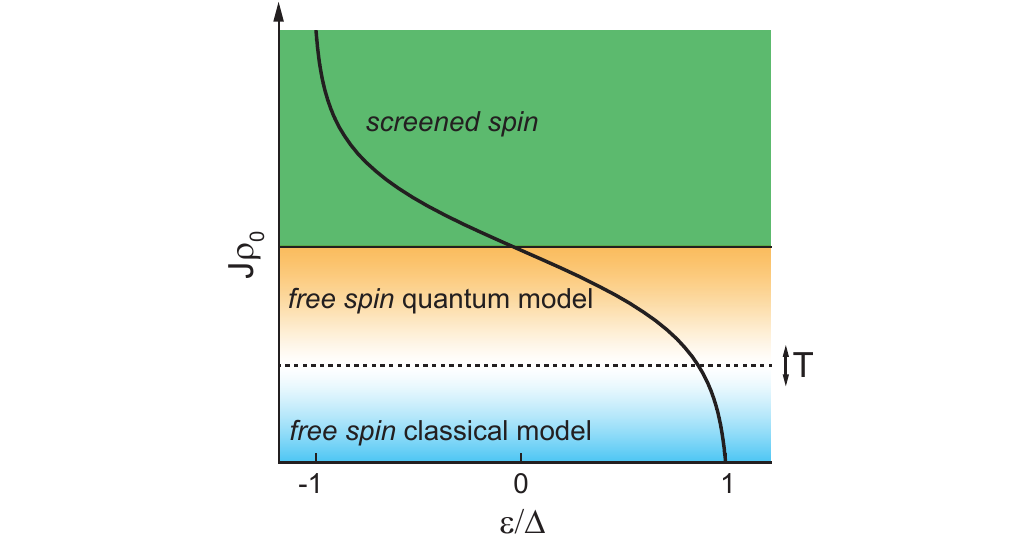} 
	\caption{Sketch of the YSR phase diagram. The transition from a classical spin model to a quantum description depends on the temperature.}
\label{Sketch:YSR}
\end{figure}

The exchange coupling strength $J$ controls the magnetic properties of an impurity on a superconductor. It determines whether the adsorbate maintains a spin and, therefore, a magnetic moment, or whether the spin is screened. Usually, the exchange coupling strength cannot be measured directly, albeit its impact is seen in Kondo resonances and YSR states. Because $J$ is responsible for both, the energy scale of the Kondo effect and the energy of the YSR states, there exists a universal relation between these phenomena \cite{Matsuura77,Satori1992,Sakai1993}. Typically, this relation is discussed at $T=0$ (or at least at $T\ll T_\mathrm{K}$), neglecting the energy scale of the experimental temperature. Though, temperature  plays an important role. In particular, it can drive the crossover from a coherent, Kondo-screened spin state at $T\ll T_\mathrm{K}$ to a state at $T\gg T_\mathrm{K}$, where a coherent many-body ground state is absent and spin-scattering processes can be described to some leading order in perturbation theory. 

In contrast to Kondo physics, the energy of YSR states is not influenced by temperature (since $\Delta\approx$\,const. for $T_{exp}\lessapprox T_c/2$). However, it depends on the energy scale of the Kondo effect. Therefore, an understanding of the regime $T\gg T_\mathrm{K}$ is necessary in order to fully capture all relevant energy scales. 
\textit{A priori}, it was not obvious that the YSR resonances can  be treated classically in this regime.
Our experiments reveal a relation to the exchange coupling $J\rho_0$, as expected intuitively, and a good agreement with the treatment in the classical limit. 
We note that a \textit{free spin} is not a sufficient condition for a classical description. The free spin regime also exists in the limit of $T\ll T_\mathrm{K}$, where the quantum mechanical description is required. Hence, we add another regime to the well-known phase diagram of magnetic impurities on superconductors (sketch in Fig.~\ref{Sketch:YSR}). Contrary to the transition between the \textit{screened spin} and \textit{free spin} state, this transition depends on temperature. Our results thus conclude with a picture of all necessary energy scales. 

This is of interest not only for the single-impurity problem as discussed here, but also for the exotic physics of magnetic, nanoscale structures on superconductors. Their theoretical description often relies on classical spin models, because of the quantum impurity model being impossible to treat with analytical methods \cite{Zitko2017}. In particular, the suggestion of Majorana bound states in magnetic impurity chains and arrays on $s$-wave superconductors has been put forward in the classical spin description \cite{Nadj2013,Pientka2013,Klinovaja2013,Nakosai2013,Braunecker2013,Vazifeh2013,Nadj2014,Kim2014,Peng2015,ruby2015}.

\section{Acknowledgments}
We thank F. von Oppen,  J.\ I. Pascual, and M. Ternes for enlightening discussions. We acknowledge funding by the Deutsche Forschungsgemeinschaft through Grant No. FR2726/4 and HE7368/2 
as well as by the European Research Council through Consolidator Grant {\it NanoSpin}.

\section{Methods}
The Pb(111) single crystal surface was thoroughly cleaned by cycles of  Ne$^+$-ion sputtering and subsequent annealing to 430\,K. MnPc was deposited from a Knudsen cell at $\approx710$\,K onto the clean Pb(111) surface held at room temperature. Subsequently,  3 to 6 Langmuir of Ammonia (\NH) were dosed onto the as-prepared sample cooled to $\approx15$\,K. In order to desorb excess ammonia, the sample was annealed to $\approx100$\,K for several hours. This procedure ensured that $\approx 90\%$ of the MnPc molecules are coordinated by a single \NH\ molecule, while the Pb(111) surface was free of \NH.

All measurements were performed in a {\sc Specs} JT-STM at a temperature of 1.1\,K if not stated differently.   
Pb-covered, superconducting tips for high resolution spectra were obtained by macroscopic indentations of a W tip into the clean Pb sample following Ref.~\cite{Franke2011}. We used only tips, which showed bulk-like superconductivity, i.e., where the gap parameter of the tip $\Delta_{tip}$ was equal to the sample's gap parameter $\Delta_{sample}$. Hence, we can write $\Delta_{tip}=\Delta_{sample}=\Delta$. In \didv spectra, the superconducting tip then shifts all energies by $\pm\Delta$ and the gap size amounts to $4\Delta$ with the quasiparticle excitation peaks on the pristine surface at $\pm 2\Delta$. See the Supplementary Information of Ref.~\cite{ruby2015} for more details on the determination of $\Delta_{tip}$.
 
Measurements of the sample  in the normal metal state were performed at 0.1\,T with a  Au-covered W tip (Fig.~\ref{Kondo:B}), or with a Pb-covered tip at 2.7\,T (Fig.~\ref{Kondo:YSR}).
Spectra of the differential conductance \didv\ were acquired with standard lock-in technique at a frequency of 912\,Hz and a root-mean-square (rms) bias modulation $V_{mod}$ as indicated in the figure captions.

\clearpage

\renewcommand{\theequation}{S\arabic{equation}}
\renewcommand{\thefigure}{S\arabic{figure}}
	\setcounter{figure}{0}


\section{{Supplementary Information}}

\subsection{Controlled \NH ligand desorption}

\begin{figure}[b]
	\centering
		\includegraphics[width=0.48\textwidth]{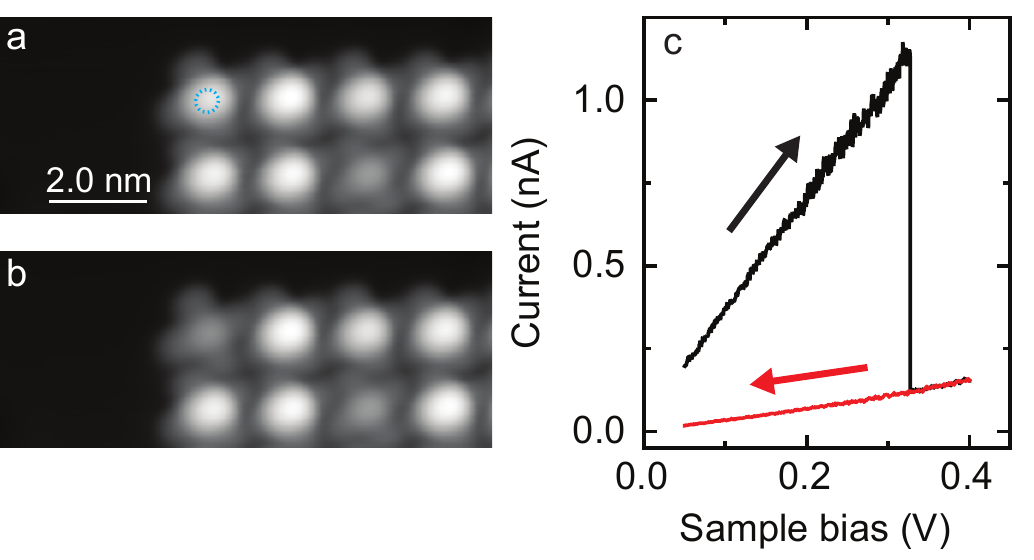}
	\caption{(a,b) Topography of a mixed island of MnPc-\NH and MnPc before (a) and after (b) controlled desorption of an \NH\ ligand. Setpoint: $50\,$mV; 200\,pA. (c) $I-V$ characteristic recorded on the position indicated with a blue circle in (a). The sudden drop in the current in the forward scan (black arrow) indicates the desorption of NH$_3$. The difference in conductance in the backward scan (red arrow) then proves the structural modification of the junction. }
\label{Sfig1}
\end{figure}

The adsorption of the neutral ligand ammonia (NH$_3$) on MnPc alters the coupling to the Pb substrate without changing the oxidation or spin state as we show in the main text. Figure~\ref{Sfig1} shows how we can remove the central \NH\ ligand from the molecule in a controlled manner without destroying the order of the molecular island. 
Figure~\ref{Sfig1}a shows a close-up of a mixed monolayer island of MnPc-\NH\ and MnPc. Next, the STM tip is placed at the position marked by a blue circle and the feedback is disabled. Then, we ramp the bias and and record the current (Fig.~\ref{Sfig1}c).
A sudden drop in the current at $\approx 0.32\,$mV indicates a change in the junction configuration. The backward sweep (red arrow) confirms an irreversible modification. Scanning the same area again (Fig.~\ref{Sfig1}b) then shows the removal of \NH\ from the target molecule and an otherwise unchanged molecular island. Spectroscopic characterization of the molecule after \NH desorption shows all features characteristic for MnPc (see below).

\subsection{Energy of YSR states in MnPc and MnPc-\NH}
\begin{figure}[b]
	\centering
		\includegraphics[width=0.48\textwidth]{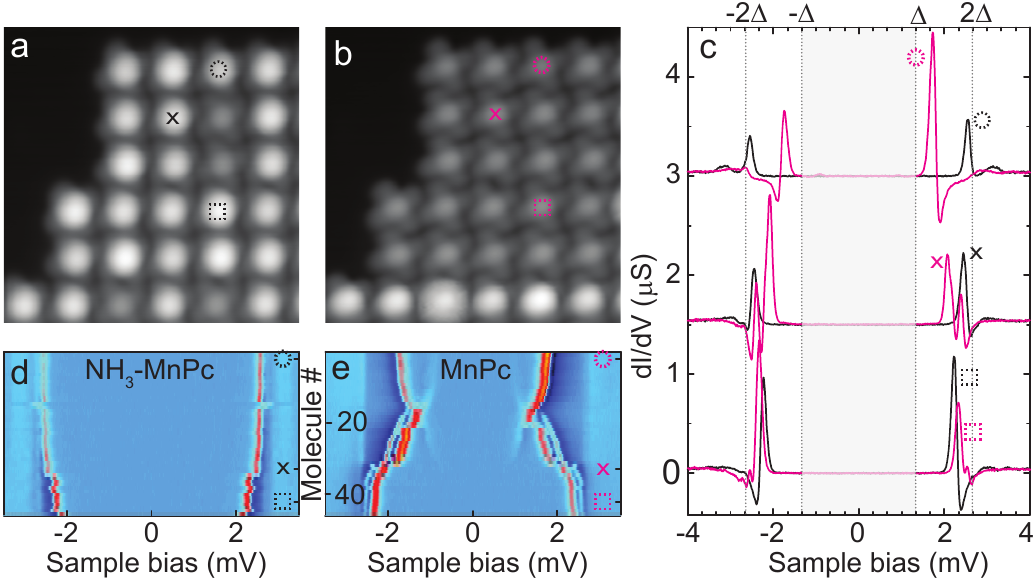}
	\caption{(a,b) Topography of a molecular monolayer island of MnPc-\NH before (a) and after (b) voltage induced desorption of some \NH ligands. Setpoint: $50\,$mV; 200\,pA. (c) three \didv excitation spectra of MnPc-\NH (black) as indicated in (a), and of the same molecules after \NH desorption (pink).  (d) 2D color plot of \didv excitation spectra of 44 MnPc-\NH molecules ordered according to the order in (e). (e) 2D color plot of \didv excitation spectra of MnPc molecules (same as in (d), but after \NH desorption). Spectra are ordered according to the energy of the YSR resonance. Spectra acquired with a superconducting Pb tip. Setpoint: $5\,$mV, 200\,pA; lock-in parameters: 912\,Hz, $15\,\mu$V$_{rms}$.}
\label{Topo:Shiba}
\end{figure}

Figure\,\ref{Topo:Shiba} shows an island of MnPc-\NH and MnPc before (a) and after (b) desorption of \NH ligands.  
To unveil the effect of the axial ligand on the exchange coupling strength between the molecules and the superconductor, we perform \didv\ spectroscopy with a superconducting tip before and after desorption of the \NH\ ligand. Figure\,\ref{Topo:Shiba}c (black curves) shows characteristic examples of \didv\ spectra on three different MnPc-NH$_3$ molecules. For all complexes, we observe a single pair of YSR excitations. While the excitation energy varies from molecule to molecule, the electron-like excitation, i.e., the resonance at positive energy, is more intense for all  MnPc-\NH molecules studied.
This is in contrast to \didv\ spectra on pristine MnPc, which were acquired after the controlled desorption of the \NH ligand (pink curves in Fig.\,\ref{Topo:Shiba}c). The YSR resonances of MnPc are split due to magnetocrystalline anisotropy~\cite{hatter2015} and show stronger variations in energy and even a shift of the main spectral weight to the negative bias side, i.e., from the electron-like to the hole-like excitations. 

For pristine MnPc, the variations of the YSR binding energy were interpreted in terms of a long-range Moir\'e pattern caused by the incommensurability of the fourfold symmetric molecular island structure with the threefold symmetric substrate. This leads to variations in the adsorption site within the molecular island and, hence, of the exchange coupling strength. As a result, a quantum  phase transition of the many-body ground state from a {\it Kondo-screened} to a {\it free-spin} ground state as indicated by the shift of the spectra weight of the YSR resonances from the positive to the negative bias side is observed~\cite{Franke2011,hatter2015}.  
\begin{figure}[b]
	\centering
		\includegraphics[width=0.48\textwidth]{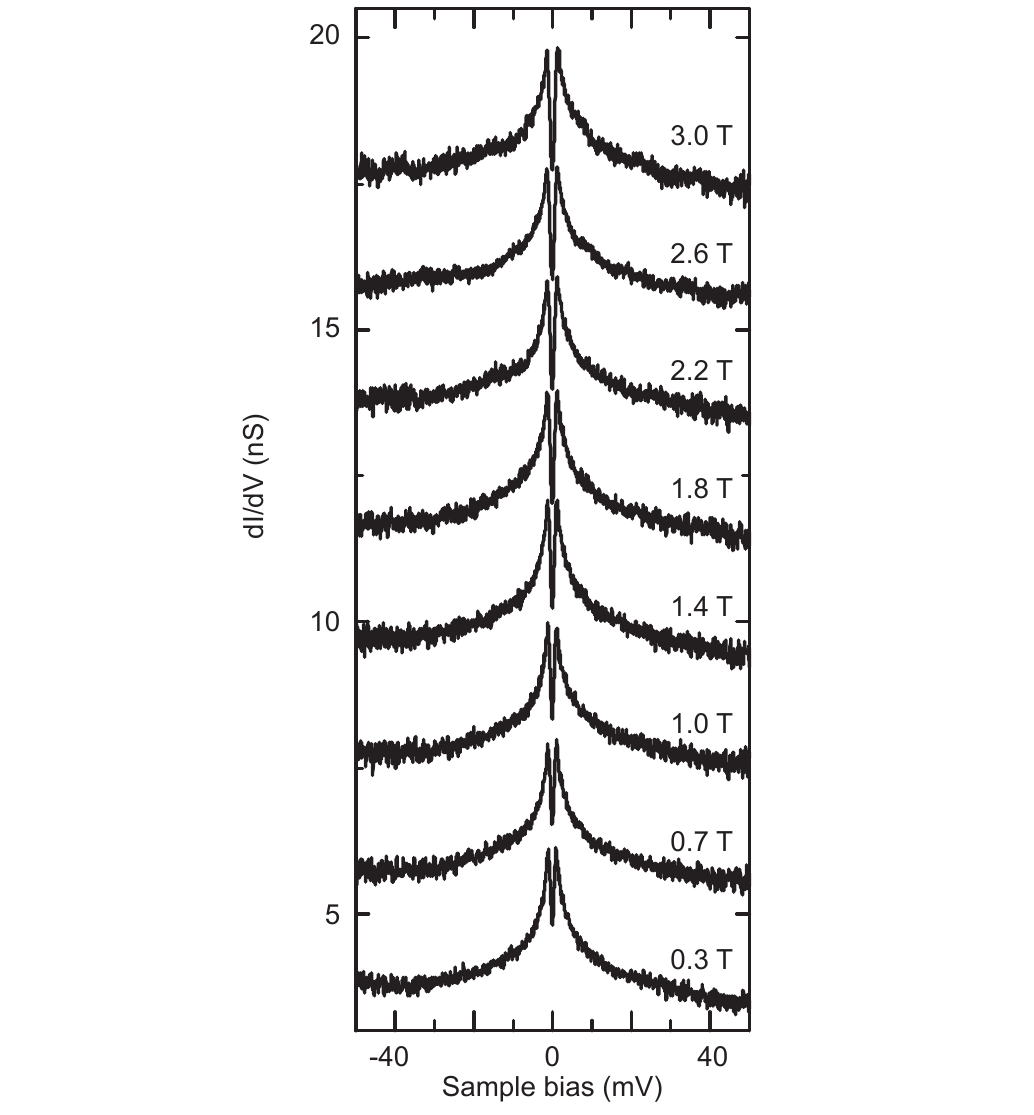}
	\caption{$B$ field evolution of the zero-energy resonance. \didv\ spectra of MnPc-\NH at 1.1\,K acquired with a normal metal Au tip. Setpoint: $U=50\,$mV, $I=200\,$pA; $U_\textrm{mod}=500\,\mu$V$_{rms}$. Same molecule as in Fig.~3 of the main text.  }
	\label{Sfig2}
\end{figure}

To gain a systematic understanding of the changes induced by the \NH ligand on the MnPc, we plot  the \didv\ spectra of 44 MnPc molecules with and without \NH ligand in pseudo-2D color plots in Fig.\,\ref{Topo:Shiba}d and \,\ref{Topo:Shiba}e, respectively. The spectra are ordered according to the YSR binding energy of the MnPc molecules without ligand. We order the spectra with decreasing energy, i.e., with increasing exchange coupling strength from top to bottom. 
The spectra of the same  molecules with \NH ligand are plotted in the same order, i.e., according to the energy of the MnPc YSR states. 
This order is then also applied to the spectra of MnPc-NH$_3$, acquired before the ligand was detached via a voltage pulse (Fig.\,\ref{Topo:Shiba}d). All YSR binding energies are shifted to more positive values (into a range not observed for MnPc). Interestingly, the order of increasing coupling strength from top to bottom is preserved.     
Hence, the \NH ligand decreased the coupling to the substrate via the surface trans effect, but does not bleach the differences in the coupling to the substrate induced by the adsorption site differences.

\subsection{Evolution of the zero-energy resonance with magnetic field}

In Fig.\,3b of the main text, we present the evolution of the symmetric steps around $E_{\mathrm F}$ in a magnetic field perpendicular to the sample surface. For completeness, Fig.\,\ref{Sfig2} presents \didv\ spectra at the same $B$ fields acquired on the same molecule with the same tip  focusing on the zero-energy resonance, i.e., in a wider energy range. While we observe that the gap-like feature at $E_{\mathrm F}$ gets more pronounced (this is because the inelastic excitation steps move to higher energy), the zero-energy  resonance appears unchanged. 

\subsection{Temperature evolution of the zero-energy resonance - fit to the strong-coupling Kondo model}

\begin{figure}[b] 
	\centering
		\includegraphics[width=0.48\textwidth]{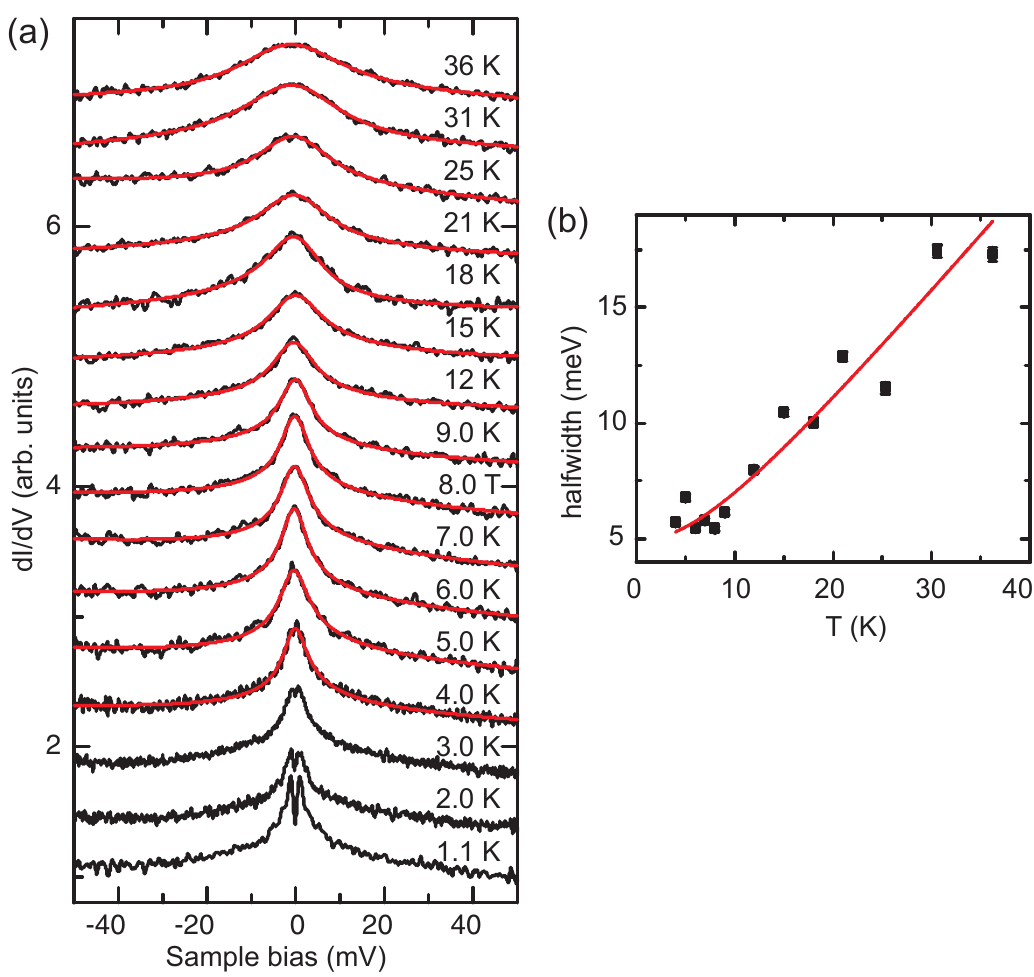} 
	\caption{(a) Temperature-dependent \didv\ spectra (same data as in Fig.~3a of the main manuscript) with a Fano-Frota fit, which accounts for temperature broadening ($50\,$mV, 200\,pA, $500\,\mu$V$_{rms}$). (b) Half width of the resonance as extracted from the Fano-Frota fit in (a) as a function of temperature. The red line is a  fit describing the temperature evolution in the strong-coupling Kondo regime as described in the text. Fit parameters: $T_\mathrm{K} = (57\pm5)\,$K; $\alpha=(3.7\pm0.2)\,\pi$. 
}
	\label{Sfig5}
\end{figure}

In the main text, we show that the zero-energy resonance is well described by perturbation theory accounting for third order  spin scattering processes.  For comparison, we show in Fig.\,\ref{Sfig5}a  fits of the temperature-dependent \didv\ spectra with a Fano-Frota function accounting for temperature broadening. This function describes the transmission for a strong-coupling Kondo impurity, i.e., for $T_\mathrm{K} \gg T$ ($T_\mathrm{K}$ is the Kondo temperature, T the experimental temperature) ~\cite{Frota1992,Pruser2011,Frank2015}. 
We extract the half width at half maximum (HWHM) from the fits and plot it as a function of $T$ (Fig.\,\ref{Sfig5}b). A fit to $\mathrm{HWHM}=\frac{1}{2}\sqrt{(\alpha k_\mathrm{B}T)^2+(2 k_\mathrm{B}T_\mathrm{K})^2}$ is shown in red ($k_\mathrm{B}$ is the Boltzmann constant). According to  Fermi liquid theory, in the strong-coupling Kondo regime $\alpha = 2\pi$ \cite{Nagaoka2002}, but the fit yields $\alpha=(3.7\pm0.2)\pi$.  This clear deviation is another indicator for the system to be in the weak-coupling Kondo regime and renders the determination of a Kondo temperature via a fit of the spectra erroneous~\cite{Zhang2013}. This interpretation agrees with the zero-field splitting of the resonance (see main text) and is in line with the observed YSR energies  of $\epsilon/\Delta \geq +0.5$, which indicate Kondo temperatures well below $\Delta$~\cite{Satori1992}. 


\begin{figure}[h] 
	\centering
		\includegraphics[width=0.48\textwidth]{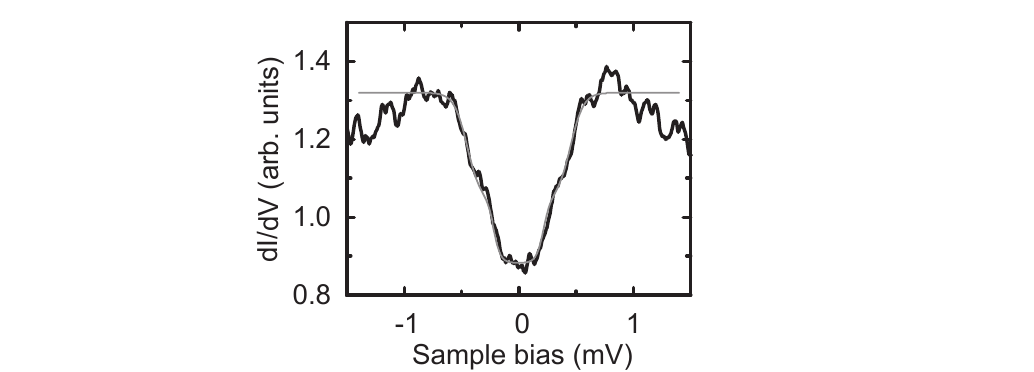} 
	\caption{\didv\ spectrum on an MnPc-\NH at $B=0.1\,$T acquired with a Au tip. The gap-like feature  shows two pairs of symmetric steps around $E_{\mathrm F}$. Setpoint: 5\,mV, 200\,pA; $U_{mod}= 50\, \mu$V$_{rms}$. The grey line presents a fit with two pairs of step functions at a sample bias of $\pm(0.22\pm0.05)$\,mV and  $\pm(0.45\pm0.05)$\,mV, respectively.}
	\label{Sfig3}
\end{figure}

\subsection{Inelastic excitations with two pairs of steps}

For most of the MnPc-\NH molecules, we observe a single pair of symmetric steps  around $E_{\mathrm F}$, which are identified as an inelastic spin excitations within the zero-field split spin $S=1$ manifold. Here, we observe two inelastic excitations which are due to a sizeable rhombicity ($E o$) and, hence, a lifting of the $M_s=\pm 1$ degeneracy. The fit with two pairs of step functions symmetric to zero yields $D=(-0.34\pm0.07)$\,meV and $E=(0.11\pm0.03)$\,meV. Because of the limited energy resolution with a metallic tip ($\approx 300\,\mu$eV at 1.1\,K), two pairs of steps are only observed for the largest $E$ anisotropies of the ensemble.

\subsection{Zeeman shift of inelastic excitations}

In Fig.\,3e of the main text, we show the $B$ field dependent energies of the inelastic spin excitation of an MnPc-\NH complex with vanishing rhombicity ($E\approx0$). A linear evolution with $B$ is observed. Here, we present the data of three different molecules with non-vanishing $E$. At  fields larger than 1\,T, the excitation energy also increases linearly with field, but below 1\,T, the increase is sublinear. This is in line with an easy-axis anisotropy ($D<0$) with some in-plane distortion ($E\neq 0$).

\begin{figure}[h] 
	\centering
		\includegraphics[width=0.48\textwidth]{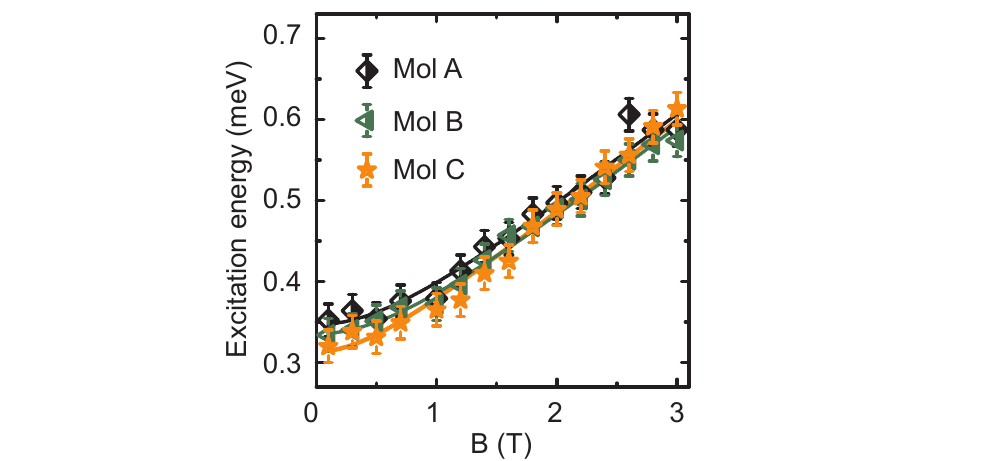} 
	\caption{Excitation energies as a function of $B$ field as extracted from a fit with symmetric step functions of the \didv\ spectra of three MnPc-\NH complexes $A$, $B$, and $C$.  The symbols are experimental data points. The full lines are fits to the Spin Hamiltonian described in the text. Fit parameters in mV: $D_a=-0.24\pm0.02$, $E_a=0.10\pm0.02$; $D_b=-0.23\pm0.01$, $E_b=0.11\pm0.01$; $D_c=-0.24\pm0.01$, $E_c=0.07\pm0.02$. }
	\label{Sfig4}
\end{figure}
We fit the data points to a Spin Hamiltonian that assumes the main anisotropy axis to be parallel to the out-of-plane field and accounts for the in-plane distortion: 
\begin{equation}
\mathbf{\hat{\cal H}}=DS_z^2+E(S_x^2-S_y^2)-g\mu_\mathrm{B} B_z S_z.
\end{equation}

 Here, $D$, $E$, $S_i$, $B_i$ , $\mu_\mathrm{B}$, and $g$ are the axial anisotropy parameter, the in-plane distortion, the projection  of the spin and the magnetic field  in  direction $i$, the Bohr magneton, and the  Land\'e g-factor, respectively. In order to keep the number of free fit parameter small we restrict $g$ to $2$. 
 \\


\begin{thebibliography}{0}%
\makeatletter
\providecommand \@ifxundefined [1]{%
 \@ifx{#1\undefined}
}%
\providecommand \@ifnum [1]{%
 \ifnum #1\expandafter \@firstoftwo
 \else \expandafter \@secondoftwo
 \fi
}%
\providecommand \@ifx [1]{%
 \ifx #1\expandafter \@firstoftwo
 \else \expandafter \@secondoftwo
 \fi
}%
\providecommand \natexlab [1]{#1}%
\providecommand \enquote  [1]{``#1''}%
\providecommand \bibnamefont  [1]{#1}%
\providecommand \bibfnamefont [1]{#1}%
\providecommand \citenamefont [1]{#1}%
\providecommand \href@noop [0]{\@secondoftwo}%
\providecommand \href [0]{\begingroup \@sanitize@url \@href}%
\providecommand \@href[1]{\@@startlink{#1}\@@href}%
\providecommand \@@href[1]{\endgroup#1\@@endlink}%
\providecommand \@sanitize@url [0]{\catcode `\\12\catcode `\$12\catcode
  `\&12\catcode `\#12\catcode `\^12\catcode `\_12\catcode `\%12\relax}%
\providecommand \@@startlink[1]{}%
\providecommand \@@endlink[0]{}%
\providecommand \url  [0]{\begingroup\@sanitize@url \@url }%
\providecommand \@url [1]{\endgroup\@href {#1}{\urlprefix }}%
\providecommand \urlprefix  [0]{URL }%
\providecommand \Eprint [0]{\href }%
\providecommand \doibase [0]{http://dx.doi.org/}%
\providecommand \selectlanguage [0]{\@gobble}%
\providecommand \bibinfo  [0]{\@secondoftwo}%
\providecommand \bibfield  [0]{\@secondoftwo}%
\providecommand \translation [1]{[#1]}%
\providecommand \BibitemOpen [0]{}%
\providecommand \bibitemStop [0]{}%
\providecommand \bibitemNoStop [0]{.\EOS\space}%
\providecommand \EOS [0]{\spacefactor3000\relax}%
\providecommand \BibitemShut  [1]{\csname bibitem#1\endcsname}%
\let\auto@bib@innerbib\@empty
\end{thebibliography}%


\begin{thebibliography}{10}
\expandafter\ifx\csname url\endcsname\relax
  \def\url#1{\texttt{#1}}\fi
\expandafter\ifx\csname urlprefix\endcsname\relax\def\urlprefix{URL }\fi
\providecommand{\bibinfo}[2]{#2}
\providecommand{\eprint}[2][]{\url{#2}}

\bibitem{Yu1965}
\bibinfo{author}{Yu, L.}
\newblock \bibinfo{title}{{Bound state in superconducors with paramagnetic
  impurities}}.
\newblock \emph{\bibinfo{journal}{Acta Physica Sinica}}
  \textbf{\bibinfo{volume}{21}}, \bibinfo{pages}{75--91}
  (\bibinfo{year}{1965}).

\bibitem{Shiba1968}
\bibinfo{author}{Shiba, H.}
\newblock \bibinfo{title}{{Classical Spins in Superconductors}}.
\newblock \emph{\bibinfo{journal}{Progress of Theoretical Physics}}
  \textbf{\bibinfo{volume}{40}}, \bibinfo{pages}{435--451}
  (\bibinfo{year}{1968}).

\bibitem{rusinov_a.i._theory_1969}
\bibinfo{author}{{Rusinov A.I.}}
\newblock \bibinfo{title}{{On the theory of gapless superconductivity in alloys
  containing paramagnetic impurities}}.
\newblock \emph{\bibinfo{journal}{Sov. Phys. JETP}}
  \textbf{\bibinfo{volume}{29}}, \bibinfo{pages}{1101--1106}
  (\bibinfo{year}{1969}).

\bibitem{yazdani_probing_1997}
\bibinfo{author}{Yazdani, A.}, \bibinfo{author}{Jones, B.~A.},
  \bibinfo{author}{Lutz, C.~P.}, \bibinfo{author}{Crommie, M.~F.} \&
  \bibinfo{author}{Eigler, D.~M.}
\newblock \bibinfo{title}{{Probing the Local Effects of Magnetic Impurities on Superconductivity}}.
\newblock \emph{\bibinfo{journal}{Science}} \textbf{\bibinfo{volume}{275}},
  \bibinfo{pages}{1767--1770} (\bibinfo{year}{1997}).

\bibitem{Ji2008}
\bibinfo{author}{Ji, S.-H.} \emph{et~al.}
\newblock \bibinfo{title}{{High-Resolution Scanning Tunneling Spectroscopy of
  Magnetic Impurity Induced Bound States in the Superconducting Gap of Pb Thin
  Films}}.
\newblock \emph{\bibinfo{journal}{Physical Review Letters}}
  \textbf{\bibinfo{volume}{100}}, \bibinfo{pages}{226801}
  (\bibinfo{year}{2008}).

\bibitem{Heinrich2017rev}
\bibinfo{author}{{Heinrich}, B.~W.}, \bibinfo{author}{{Pascual}, J.~I.} \&
  \bibinfo{author}{{Franke}, K.~J.}
\newblock \bibinfo{title}{{Single magnetic adsorbates on s-wave
  superconductors}}.
\newblock \eprint{Preprint at https://arxiv.org/abs/1705.03672} (\bibinfo{year}{2017}).

\bibitem{Matsuura77}
\bibinfo{author}{Matsuura, T.}
\newblock \bibinfo{title}{{The Effects of Impurities on Superconductors with
  Kondo Effect}}.
\newblock \emph{\bibinfo{journal}{Progress of Theoretical Physics}}
  \textbf{\bibinfo{volume}{57}}, \bibinfo{pages}{1823--1835}
  (\bibinfo{year}{1977}).

\bibitem{Satori1992}
\bibinfo{author}{Satori, K.}, \bibinfo{author}{Shiba, H.},
  \bibinfo{author}{Sakai, O.} \& \bibinfo{author}{Shimizu, Y.}
\newblock \bibinfo{title}{{Numerical renormalization group study of magnetic
  impurities in superconductors}}.
\newblock \emph{\bibinfo{journal}{Journal of the Physical Society of Japan}}
  \textbf{\bibinfo{volume}{61}}, \bibinfo{pages}{3239--3254}
  (\bibinfo{year}{1992}).

\bibitem{Sakai1993}
\bibinfo{author}{Sakai, O.}, \bibinfo{author}{Shimizu, Y.},
  \bibinfo{author}{Shiba, H.} \& \bibinfo{author}{Satori, K.}
\newblock \bibinfo{title}{{Numerical Renormalization Group Study of Magnetic Impurities in Superconductors. II. Dynamical Excitation Spectra and Spatial Variation of the Order Parameter}}.
\newblock \emph{\bibinfo{journal}{Journal of the Physical Society of Japan}}
  \textbf{\bibinfo{volume}{62}}, \bibinfo{pages}{3181--3197}
  (\bibinfo{year}{1993}).

\bibitem{Deacon2010b}
\bibinfo{author}{Deacon, R.~S.} \emph{et~al.}
\newblock \bibinfo{title}{{Tunneling Spectroscopy of Andreev Energy Levels in a
  Quantum Dot Coupled to a Superconductor}}.
\newblock \emph{\bibinfo{journal}{Physical Review Letters}}
  \textbf{\bibinfo{volume}{104}}, \bibinfo{pages}{076805}
  (\bibinfo{year}{2010}).

\bibitem{Franke2011}
\bibinfo{author}{Franke, K.~J.}, \bibinfo{author}{Schulze, G.} \&
  \bibinfo{author}{Pascual, J.~I.}
\newblock \bibinfo{title}{{Competition of superconducting phenomena and Kondo screening at the nanoscale}}.
\newblock \emph{\bibinfo{journal}{Science}} \textbf{\bibinfo{volume}{332}},
  \bibinfo{pages}{940--945} (\bibinfo{year}{2011}).


\bibitem{Lee2014}
\bibinfo{author}{Lee, E. J.~H.} \emph{et~al.}
\newblock \bibinfo{title}{{Spin-resolved Andreev levels and parity crossings in
  hybrid superconductor-semiconductor nanostructures.}}
\newblock \emph{\bibinfo{journal}{Nature Nanotechnology}}
  \textbf{\bibinfo{volume}{9}}, \bibinfo{pages}{79--84} (\bibinfo{year}{2014}).

\bibitem{Jellinggaard2016}
\bibinfo{author}{Jellinggaard, A.}, \bibinfo{author}{Grove-Rasmussen, K.},
  \bibinfo{author}{Madsen, M.~H.} \& \bibinfo{author}{Nyg{\aa}rd, J.}
\newblock \bibinfo{title}{{Tuning Yu-Shiba-Rusinov states in a quantum dot}}.
\newblock \emph{\bibinfo{journal}{Physical Review B}}
  \textbf{\bibinfo{volume}{94}}, \bibinfo{pages}{064520}
  (\bibinfo{year}{2016}).

\bibitem{Lee2017}
\bibinfo{author}{Lee, E. J.~H.} \emph{et~al.}
\newblock \bibinfo{title}{Scaling of subgap excitations in a
  superconductor-semiconductor nanowire quantum dot}.
\newblock \emph{\bibinfo{journal}{Physical Review B}} \textbf{\bibinfo{volume}{95}},
  \bibinfo{pages}{180502} (\bibinfo{year}{2017}).

\bibitem{Zhang2013}
\bibinfo{author}{Zhang, Y.-H.} \emph{et~al.}
\newblock \bibinfo{title}{{Temperature and magnetic field dependence of a Kondo
  system in the weak coupling regime.}}
\newblock \emph{\bibinfo{journal}{Nature Communications}}
  \textbf{\bibinfo{volume}{4}}, \bibinfo{pages}{2110} (\bibinfo{year}{2013}).

\bibitem{Ternes2016}
\bibinfo{author}{Ternes, M.}
\newblock \bibinfo{title}{Spin excitations and correlations in scanning
  tunneling spectroscopy}.
\newblock \emph{\bibinfo{journal}{New Journal of Physics}}
  \textbf{\bibinfo{volume}{17}}, \bibinfo{pages}{063016}
  (\bibinfo{year}{2015}).

\bibitem{Shuai2010}
\bibinfo{author}{Shuai-Hua, J.} \emph{et~al.}
\newblock \bibinfo{title}{{Kondo Effect in Self-Assembled Manganese
  Phthalocyanine Monolayer on Pb Islands Kondo Effect in Self-Assembled
  Manganese Phthalocyanine Monolayer on Pb}}.
\newblock \emph{\bibinfo{journal}{Chinese Physics Letters}}
  \textbf{\bibinfo{volume}{27}}, \bibinfo{pages}{087202}
  (\bibinfo{year}{2010}).

\bibitem{hatter2015}
\bibinfo{author}{Hatter, N.}, \bibinfo{author}{Heinrich, B.~W.},
  \bibinfo{author}{Ruby, M.}, \bibinfo{author}{Pascual, J.~I.} \&
  \bibinfo{author}{Franke, K.~J.}
\newblock \bibinfo{title}{{Magnetic anisotropy in Shiba bound states across a
  quantum phase transition}}.
\newblock \emph{\bibinfo{journal}{Nature Communications}}
  \textbf{\bibinfo{volume}{6}}, \bibinfo{pages}{8988} (\bibinfo{year}{2015}).

\bibitem{Flechtner2007}
\bibinfo{author}{Flechtner, K.}, \bibinfo{author}{Kretschmann, A.},
  \bibinfo{author}{Steinr{\"{u}}ck, H.-P.} \& \bibinfo{author}{Gottfried, J.~M.}
\newblock \bibinfo{title}{{NO-induced Reversible Switching of the Electronic
  Interaction between a Porphyrin-Coordinated Cobalt Ion and a Silver
  Surface}}.
\newblock \emph{\bibinfo{journal}{J. Am. Chem. Soc.}}
  \textbf{\bibinfo{volume}{129}}, \bibinfo{pages}{12110 -- 12111}
  (\bibinfo{year}{2007}).

\bibitem{Hieringer2011}
\bibinfo{author}{Hieringer, W.} \emph{et~al.}
\newblock \bibinfo{title}{{The surface trans effect: influence of axial ligands
  on the surface chemical bonds of adsorbed metalloporphyrins.}}
\newblock \emph{\bibinfo{journal}{Journal of the American Chemical Society}}
  \textbf{\bibinfo{volume}{133}}, \bibinfo{pages}{6206--22}
  (\bibinfo{year}{2011}).

\bibitem{Zitko2011}
\bibinfo{author}{{\v{Z}}itko, R.}, \bibinfo{author}{Bodensiek, O.} \&
  \bibinfo{author}{Pruschke, T.}
\newblock \bibinfo{title}{{Effects of magnetic anisotropy on the subgap
  excitations induced by quantum impurities in a superconducting host}}.
\newblock \emph{\bibinfo{journal}{Physical Review B}}
  \textbf{\bibinfo{volume}{83}}, \bibinfo{pages}{054512}
  (\bibinfo{year}{2011}).

\bibitem{isvoranu_ammonia_2011}
\bibinfo{author}{Isvoranu, C.} \emph{et~al.}
\newblock \bibinfo{title}{{Ammonia adsorption on iron phthalocyanine on
  Au(111): Influence on adsorbate–substrate coupling and
  molecular spin}}.
\newblock \emph{\bibinfo{journal}{The Journal of Chemical Physics}}
  \textbf{\bibinfo{volume}{134}} (\bibinfo{year}{2011}).

\bibitem{note1}
\bibinfo{title}{The Zeeman energy at 0.1\,T is $\sim 11\,\mu$eV (for $g\approx 2$), which is ten times smaller than the thermal resolution at 1.1\,K.}

\bibitem{note2}
\bibinfo{title}{We also attempted a fit with a Fano-Frota function including temperature broadening in the Supplementary Information. This analysis corroborates that the system is not in the strong-coupling Kondo regime.}

\bibitem{Appelbaum1966}
\bibinfo{author}{Appelbaum, J.}
\newblock \bibinfo{title}{"$s\ensuremath{-}d$" exchange model of zero-bias
  tunneling anomalies}.
\newblock \emph{\bibinfo{journal}{Physical Review Letters}}
  \textbf{\bibinfo{volume}{17}}, \bibinfo{pages}{91--95}
  (\bibinfo{year}{1966}).

\bibitem{Anderson1966}
\bibinfo{author}{Anderson, P.~W.}
\newblock \bibinfo{title}{Localized magnetic states and Fermi-surface anomalies
  in tunneling}.
\newblock \emph{\bibinfo{journal}{Physical Review Letters}}
  \textbf{\bibinfo{volume}{17}}, \bibinfo{pages}{95--97}
  (\bibinfo{year}{1966}).

\bibitem{Appelbaum1967}
\bibinfo{author}{Appelbaum, J.~A.}
\newblock \bibinfo{title}{Exchange model of zero-bias tunneling anomalies}.
\newblock \emph{\bibinfo{journal}{Physical Review }} \textbf{\bibinfo{volume}{154}},
  \bibinfo{pages}{633--643} (\bibinfo{year}{1967}).


\bibitem{note3}
\bibinfo{title} {A field of $B=2.7$\,T is necessary in order to quench superconductivity also in the tip, which has an increased critical field because of its finite size.}

\bibitem{Zitko2017}
\bibinfo{author}{{\v{Z}}itko, R.}
\newblock \bibinfo{title}{{Quantum impurity models for magnetic adsorbates on superconducting surfaces}}.
\newblock \emph{\bibinfo{journal}{Physica B}}
\newblock \urlprefix\url{http://dx.doi.org/10.1016/jphysb.2017.08.019}.
  (\bibinfo{year}{2017}).



\bibitem{Nadj2013}
\bibinfo{author}{Nadj-Perge, S.}, \bibinfo{author}{Drozdov, I.~K.},
  \bibinfo{author}{Bernevig, B.~A.} \& \bibinfo{author}{Yazdani, A.}
\newblock \bibinfo{title}{Proposal for realizing Majorana Fermions in chains of
  magnetic atoms on a superconductor}.
\newblock \emph{\bibinfo{journal}{Physical Review B}} \textbf{\bibinfo{volume}{88}},
  \bibinfo{pages}{020407} (\bibinfo{year}{2013}).

\bibitem{Pientka2013}
\bibinfo{author}{Pientka, F.}, \bibinfo{author}{Glazman, L.~I.} \&
  \bibinfo{author}{von Oppen, F.}
\newblock \bibinfo{title}{Topological superconducting phase in helical Shiba
  chains}.
\newblock \emph{\bibinfo{journal}{Physical Review B}} \textbf{\bibinfo{volume}{88}},
  \bibinfo{pages}{155420} (\bibinfo{year}{2013}).

\bibitem{Klinovaja2013}
\bibinfo{author}{Klinovaja, J.}, \bibinfo{author}{Stano, P.},
  \bibinfo{author}{Yazdani, A.} \& \bibinfo{author}{Loss, D.}
\newblock \bibinfo{title}{Topological Superconductivity and Majorana Fermions
  in RKKY systems}.
\newblock \emph{\bibinfo{journal}{Physical Review Letters}}
  \textbf{\bibinfo{volume}{111}}, \bibinfo{pages}{186805}
  (\bibinfo{year}{2013}).

\bibitem{Nakosai2013}
\bibinfo{author}{Nakosai, S.}, \bibinfo{author}{Tanaka, Y.} \&
  \bibinfo{author}{Nagaosa, N.}
\newblock \bibinfo{title}{Two-dimensional $p$-wave superconducting states with
  magnetic moments on a conventional $s$-wave superconductor}.
\newblock \emph{\bibinfo{journal}{Physical Review B}} \textbf{\bibinfo{volume}{88}},
  \bibinfo{pages}{180503} (\bibinfo{year}{2013}).

\bibitem{Braunecker2013}
\bibinfo{author}{Braunecker, B.} \& \bibinfo{author}{Simon, P.}
\newblock \bibinfo{title}{Interplay between classical magnetic moments and
  superconductivity in quantum one-dimensional conductors: Toward a
  self-sustained topological Majorana phase}.
\newblock \emph{\bibinfo{journal}{Physical Review Letters}}
  \textbf{\bibinfo{volume}{111}}, \bibinfo{pages}{147202}
  (\bibinfo{year}{2013}).

\bibitem{Vazifeh2013}
\bibinfo{author}{Vazifeh, M.~M.} \& \bibinfo{author}{Franz, M.}
\newblock \bibinfo{title}{Self-organized topological state with Majorana
  Fermions}.
\newblock \emph{\bibinfo{journal}{Physical Review Letters}}
  \textbf{\bibinfo{volume}{111}}, \bibinfo{pages}{206802}
  (\bibinfo{year}{2013}).

\bibitem{Nadj2014}
\bibinfo{author}{Nadj-Perge, S.} \emph{et~al.}
\newblock \bibinfo{title}{{Observation of Majorana Fermions in ferromagnetic
  atomic chains on a superconductor}}.
\newblock \emph{\bibinfo{journal}{Science}} \textbf{\bibinfo{volume}{346}},
  \bibinfo{pages}{602--607} (\bibinfo{year}{2014}).

\bibitem{Kim2014}
\bibinfo{author}{Kim, Y.}, \bibinfo{author}{Cheng, M.}, \bibinfo{author}{Bauer,
  B.}, \bibinfo{author}{Lutchyn, R.~M.} \& \bibinfo{author}{Das~Sarma, S.}
\newblock \bibinfo{title}{Helical order in one-dimensional magnetic atom chains
  and possible emergence of Majorana bound states}.
\newblock \emph{\bibinfo{journal}{Physical Review B}} \textbf{\bibinfo{volume}{90}},
  \bibinfo{pages}{060401} (\bibinfo{year}{2014}).

\bibitem{Peng2015}
\bibinfo{author}{Peng, Y.}, \bibinfo{author}{Pientka, F.},
  \bibinfo{author}{Glazman, L.~I.} \& \bibinfo{author}{von Oppen, F.}
\newblock \bibinfo{title}{Strong localization of Majorana end states in chains
  of magnetic adatoms}.
\newblock \emph{\bibinfo{journal}{Physical Review Letters}}
  \textbf{\bibinfo{volume}{114}}, \bibinfo{pages}{106801}
  (\bibinfo{year}{2015}).

\bibitem{ruby2015}
\bibinfo{author}{Ruby, M.}, \bibinfo{author}{Pientka, F.},
  \bibinfo{author}{Peng, Y.} 
  \bibinfo{author}{von Oppen, F.}
  \bibinfo{author}{Heinrich, B. W.}
\& \bibinfo{author}{Franke, K. J.}
\newblock \bibinfo{title}{	End states and subgap structure in proximity-coupled chains of magnetic adatoms}.
\newblock \emph{\bibinfo{journal}{Physical Review Letters}}
  \textbf{\bibinfo{volume}{115}}, \bibinfo{pages}{197204}
  (\bibinfo{year}{2015}).



\end{thebibliography}

\begin{thebibliography}{8}%
\makeatletter
\providecommand \@ifxundefined [1]{%
 \@ifx{#1\undefined}
}%
\providecommand \@ifnum [1]{%
 \ifnum #1\expandafter \@firstoftwo
 \else \expandafter \@secondoftwo
 \fi
}%
\providecommand \@ifx [1]{%
 \ifx #1\expandafter \@firstoftwo
 \else \expandafter \@secondoftwo
 \fi
}%
\providecommand \natexlab [1]{#1}%
\providecommand \enquote  [1]{``#1''}%
\providecommand \bibnamefont  [1]{#1}%
\providecommand \bibfnamefont [1]{#1}%
\providecommand \citenamefont [1]{#1}%
\providecommand \href@noop [0]{\@secondoftwo}%
\providecommand \href [0]{\begingroup \@sanitize@url \@href}%
\providecommand \@href[1]{\@@startlink{#1}\@@href}%
\providecommand \@@href[1]{\endgroup#1\@@endlink}%
\providecommand \@sanitize@url [0]{\catcode `\\12\catcode `\$12\catcode
  `\&12\catcode `\#12\catcode `\^12\catcode `\_12\catcode `\%12\relax}%
\providecommand \@@startlink[1]{}%
\providecommand \@@endlink[0]{}%
\providecommand \url  [0]{\begingroup\@sanitize@url \@url }%
\providecommand \@url [1]{\endgroup\@href {#1}{\urlprefix }}%
\providecommand \urlprefix  [0]{URL }%
\providecommand \Eprint [0]{\href }%
\providecommand \doibase [0]{http://dx.doi.org/}%
\providecommand \selectlanguage [0]{\@gobble}%
\providecommand \bibinfo  [0]{\@secondoftwo}%
\providecommand \bibfield  [0]{\@secondoftwo}%
\providecommand \translation [1]{[#1]}%
\providecommand \BibitemOpen [0]{}%
\providecommand \bibitemStop [0]{}%
\providecommand \bibitemNoStop [0]{.\EOS\space}%
\providecommand \EOS [0]{\spacefactor3000\relax}%
\providecommand \BibitemShut  [1]{\csname bibitem#1\endcsname}%
\let\auto@bib@innerbib\@empty
\bibitem [{\citenamefont {Hatter}\ \emph {et~al.}(2015)\citenamefont {Hatter},
  \citenamefont {Heinrich}, \citenamefont {Ruby}, \citenamefont {Pascual},\
  and\ \citenamefont {Franke}}]{hatter2015}%
  \BibitemOpen
  \bibfield  {author} {\bibinfo {author} {\bibfnamefont {N.}~\bibnamefont
  {Hatter}}, \bibinfo {author} {\bibfnamefont {B.~W.}\ \bibnamefont
  {Heinrich}}, \bibinfo {author} {\bibfnamefont {M.}~\bibnamefont {Ruby}},
  \bibinfo {author} {\bibfnamefont {J.~I.}\ \bibnamefont {Pascual}}, \ and\
  \bibinfo {author} {\bibfnamefont {K.~J.}\ \bibnamefont {Franke}},\ }\href
  {\doibase 10.1038/ncomms9988} {\bibfield  {journal} {\bibinfo  {journal}
  {Nature Communications}\ }\textbf {\bibinfo {volume} {6}},\ \bibinfo {pages}
  {8988} (\bibinfo {year} {2015})}\BibitemShut {NoStop}%
\bibitem [{\citenamefont {Franke}\ \emph {et~al.}(2011)\citenamefont {Franke},
  \citenamefont {Schulze},\ and\ \citenamefont {Pascual}}]{Franke2011}%
  \BibitemOpen
  \bibfield  {author} {\bibinfo {author} {\bibfnamefont {K.~J.}\ \bibnamefont
  {Franke}}, \bibinfo {author} {\bibfnamefont {G.}~\bibnamefont {Schulze}}, \
  and\ \bibinfo {author} {\bibfnamefont {J.~I.}\ \bibnamefont {Pascual}},\
  }\href {\doibase 10.1126/science.1202204} {\bibfield  {journal} {\bibinfo
  {journal} {Science}\ }\textbf {\bibinfo {volume} {332}},\ \bibinfo {pages}
  {940} (\bibinfo {year} {2011})}\BibitemShut {NoStop}%
\bibitem [{\citenamefont {Frota}(1992)}]{Frota1992}%
  \BibitemOpen
  \bibfield  {author} {\bibinfo {author} {\bibfnamefont {H.~O.}\ \bibnamefont
  {Frota}},\ }\href@noop {} {\bibfield  {journal} {\bibinfo  {journal} {Phys.
  Rev. B}\ }\textbf {\bibinfo {volume} {45}},\ \bibinfo {pages} {1096}
  (\bibinfo {year} {1992})}\BibitemShut {NoStop}%
\bibitem [{\citenamefont {Pr{\"{u}}ser}\ \emph {et~al.}(2011)\citenamefont
  {Pr{\"{u}}ser}, \citenamefont {Wenderoth}, \citenamefont {Dargel},
  \citenamefont {Weismann}, \citenamefont {Peters}, \citenamefont {Pruschke},\
  and\ \citenamefont {Ulbrich}}]{Pruser2011}%
  \BibitemOpen
  \bibfield  {author} {\bibinfo {author} {\bibfnamefont {H.}~\bibnamefont
  {Pr{\"{u}}ser}}, \bibinfo {author} {\bibfnamefont {M.}~\bibnamefont
  {Wenderoth}}, \bibinfo {author} {\bibfnamefont {P.~E.}\ \bibnamefont
  {Dargel}}, \bibinfo {author} {\bibfnamefont {A.}~\bibnamefont {Weismann}},
  \bibinfo {author} {\bibfnamefont {R.}~\bibnamefont {Peters}}, \bibinfo
  {author} {\bibfnamefont {T.}~\bibnamefont {Pruschke}}, \ and\ \bibinfo
  {author} {\bibfnamefont {R.~G.}\ \bibnamefont {Ulbrich}},\ }\href {\doibase
  10.1038/nphys1876} {\bibfield  {journal} {\bibinfo  {journal} {Nature
  Physics}\ }\textbf {\bibinfo {volume} {7}},\ \bibinfo {pages} {203} (\bibinfo
  {year} {2011})}\BibitemShut {NoStop}%
\bibitem [{\citenamefont {Frank}\ and\ \citenamefont
  {Jacob}(2015)}]{Frank2015}%
  \BibitemOpen
  \bibfield  {author} {\bibinfo {author} {\bibfnamefont {S.}~\bibnamefont
  {Frank}}\ and\ \bibinfo {author} {\bibfnamefont {D.}~\bibnamefont {Jacob}},\
  }\href {\doibase 10.1103/PhysRevB.92.235127} {\bibfield  {journal} {\bibinfo
  {journal} {Phys. Rev. B}\ }\textbf {\bibinfo {volume} {92}},\ \bibinfo
  {pages} {235127} (\bibinfo {year} {2015})}\BibitemShut {NoStop}%
\bibitem [{\citenamefont {Nagaoka}\ \emph {et~al.}(2002)\citenamefont
  {Nagaoka}, \citenamefont {Jamneala}, \citenamefont {Grobis},\ and\
  \citenamefont {Crommie}}]{Nagaoka2002}%
  \BibitemOpen
  \bibfield  {author} {\bibinfo {author} {\bibfnamefont {K.}~\bibnamefont
  {Nagaoka}}, \bibinfo {author} {\bibfnamefont {T.}~\bibnamefont {Jamneala}},
  \bibinfo {author} {\bibfnamefont {M.}~\bibnamefont {Grobis}}, \ and\ \bibinfo
  {author} {\bibfnamefont {M.~F.}\ \bibnamefont {Crommie}},\ }\href {\doibase
  10.1103/PhysRevLett.88.077205} {\bibfield  {journal} {\bibinfo  {journal}
  {Phys. Rev. Lett.}\ }\textbf {\bibinfo {volume} {88}},\ \bibinfo {pages}
  {077205} (\bibinfo {year} {2002})}\BibitemShut {NoStop}%
\bibitem [{\citenamefont {Zhang}\ \emph {et~al.}(2013)\citenamefont {Zhang},
  \citenamefont {Kahle}, \citenamefont {Herden}, \citenamefont {Stroh},
  \citenamefont {Mayor}, \citenamefont {Schlickum}, \citenamefont {Ternes},
  \citenamefont {Wahl},\ and\ \citenamefont {Kern}}]{Zhang2013}%
  \BibitemOpen
  \bibfield  {author} {\bibinfo {author} {\bibfnamefont {Y.-H.}\ \bibnamefont
  {Zhang}}, \bibinfo {author} {\bibfnamefont {S.}~\bibnamefont {Kahle}},
  \bibinfo {author} {\bibfnamefont {T.}~\bibnamefont {Herden}}, \bibinfo
  {author} {\bibfnamefont {C.}~\bibnamefont {Stroh}}, \bibinfo {author}
  {\bibfnamefont {M.}~\bibnamefont {Mayor}}, \bibinfo {author} {\bibfnamefont
  {U.}~\bibnamefont {Schlickum}}, \bibinfo {author} {\bibfnamefont
  {M.}~\bibnamefont {Ternes}}, \bibinfo {author} {\bibfnamefont
  {P.}~\bibnamefont {Wahl}}, \ and\ \bibinfo {author} {\bibfnamefont
  {K.}~\bibnamefont {Kern}},\ }\href {\doibase 10.1038/ncomms3110} {\bibfield
  {journal} {\bibinfo  {journal} {Nature Communications}\ }\textbf {\bibinfo
  {volume} {4}},\ \bibinfo {pages} {2110} (\bibinfo {year} {2013})}\BibitemShut
  {NoStop}%
\bibitem [{\citenamefont {Satori}\ \emph {et~al.}(1992)\citenamefont {Satori},
  \citenamefont {Shiba}, \citenamefont {Sakai},\ and\ \citenamefont
  {Shimizu}}]{Satori1992}%
  \BibitemOpen
  \bibfield  {author} {\bibinfo {author} {\bibfnamefont {K.}~\bibnamefont
  {Satori}}, \bibinfo {author} {\bibfnamefont {H.}~\bibnamefont {Shiba}},
  \bibinfo {author} {\bibfnamefont {O.}~\bibnamefont {Sakai}}, \ and\ \bibinfo
  {author} {\bibfnamefont {Y.}~\bibnamefont {Shimizu}},\ }\href {\doibase
  10.1143/JPSJ.61.3239} {\bibfield  {journal} {\bibinfo  {journal} {Journal of
  the Physical Society of Japan}\ }\textbf {\bibinfo {volume} {61}},\ \bibinfo
  {pages} {3239} (\bibinfo {year} {1992})}\BibitemShut {NoStop}%
\end{thebibliography}

%

\end{document}